\documentclass[runningheads]{llncs}

\usepackage{orcidlink}
\usepackage[T1]{fontenc}
\usepackage{microtype}
\usepackage{amsmath}
\usepackage{graphicx}
\usepackage{booktabs}
\usepackage{array}
\usepackage{caption}
\usepackage{xcolor}
\usepackage{hyperref}
\usepackage{float}
\usepackage{tabularx}
\hypersetup{colorlinks=true, linkcolor=black, citecolor=blue, urlcolor=blue}

\graphicspath{{../../results/figures_pub/}{../../results/figures/}}

\begin{document}

\title{Does Mixture-of-Experts Actually Help Inference on Consumer and Edge Hardware? An Empirical Study}
\titlerunning{Does MoE Help Inference on Consumer and Edge Hardware?}

\author{Alfarizy Alfarizy\inst{2} \and
Hung Truong Thanh Nguyen\inst{2}\orcidlink{0000−0002−6750−9536} \and
Ren\'e Richard\inst{2} \and
Roozbeh Razavi-Far\inst{1} \and
Hung Cao\inst{2}\orcidlink{0000−0002−0788−4377}}

\authorrunning{A. Alfarizy et al.}

\institute{TSAI Lab, University of New Brunswick, Fredericton, NB, Canada \and Analytics Everywhere Lab, University of New Brunswick, Fredericton, NB, Canada\\
\email{alfarizy.alfarizy@unb.ca}}

\maketitle

\begin{abstract}
Mixture-of-Experts (MoE) language models are often described as ideal for resource-constrained inference. Each token activates only a small subset of experts, so the per-token compute cost, in floating-point operations (FLOPs), resembles that of a much smaller dense model. Whether that FLOP advantage survives in practice is far less clear. We ask whether MoE models actually run faster and cheaper than comparable dense models on consumer-grade and edge hardware. We benchmark OLMoE-1B-7B (1.3\,B active of 6.9\,B total) against three dense baselines on an Apple M2 Pro and an NVIDIA Jetson Orin Nano 8\,GB through \texttt{llama.cpp}, measuring throughput, memory, and on-device energy. The answer is device-dependent: OLMoE's active-parameter advantage is only partly realised on the laptop ($\approx$10\,\% behind the same-active Llama-3.2-1B) and erodes on the edge device ($\approx$31\,\% behind, at 2.1$\times$ the energy per token, with peak memory at the 8\,GB ceiling). Patching \texttt{llama.cpp} to time the decode graph node-by-node shows routing accounts for under 9\,\% of MoE-block compute on the cleaner edge backend, so the gap reflects total-parameter memory footprint, expert dispatch, and KV-cache pressure rather than routing. The implication is that on bandwidth-bound edge hardware, inference cost tracks total parameters, not active ones, and sparse activation does not buy back what the device is constrained on. These findings are bounded to one MoE model at this parameter scale and two devices, and we release the full measurement harness and per-run data. Our implementation is available at: \url{https://github.com/Analytics-Everywhere-Lab/edge-moe}.

\end{abstract}

\section{Introduction}

Large language models (LLMs) are no longer evaluated only in the datacenter. Privacy, latency, cost, and the simple wish to run something offline keep pushing inference onto consumer laptops, developer boards, and phone-class hardware~\cite{laskaridis2024melt,wang2024comprehensivesurveysmalllanguage,lu2024slm}. Mixture-of-Experts (MoE) architectures~\cite{shazeer2017outrageously,fedus2022switch} are often described as a good fit for that setting because each token is routed through only a small subset of expert sub-networks. The per-token floating-point cost therefore looks similar to that of a much smaller dense model. If that lower compute cost translates cleanly into lower wall-clock time and lower energy on widely available consumer hardware, MoE is a practical advantage in the edge environment. However, whether the translation actually happens is much less obvious.

This paper addresses one thread that is explicitly discussed in the mobile-LLM measurement literature. The MELT measurement study~\cite{laskaridis2024melt} notes that ``Mixture-of-Experts focusing on using subsets of weights based on the input at hand. However, these remain difficult to deploy on device, due to their memory and storage requirements.'' That sentence frames MoE-on-device as an open question. The complementary MoE-CAP study~\cite{jiang2025moecap} introduces sparsity-aware metrics (S-MBU, S-MFU) for MoE systems, but its end-to-end measurements are run on datacenter GPUs, and its edge analysis is theoretical bandwidth modelling rather than direct measurement. To our knowledge, MELT-style measurement of MoE models on real consumer and edge hardware has not been reported.

We therefore ask a practical question: do MoE language models actually deliver an inference-efficiency advantage over dense models when you run them on consumer (laptop-class) and edge (Jetson-class) hardware? We decompose this into four research questions. \textbf{RQ1} and \textbf{RQ2} ask how MoE generation throughput and peak memory compare with dense baselines on the active-parameter and memory-footprint axes. \textbf{RQ3} asks how prompt length shifts the trade-off. \textbf{RQ4}, reported only on Jetson, asks how many joules per generated token each model costs. We explain the Jetson-only energy scope in Section~\ref{sec:methodology}.

We investigate these questions empirically using direct measurement. The inference backend (\texttt{llama.cpp}), the quantization (\texttt{Q4\_K\_M}), the maximum generation length (256 tokens), and the prompt set (twelve prompts spanning six categories) are held constant. The MoE under test is OLMoE-1B-7B-0924-Instruct~\cite{muennighoff2024olmoe} with 1.3\,B active out of 6.9\,B total parameters. The three dense baselines, Llama-3.2-1B, Qwen2.5-1.5B, and Gemma-2-2B, are picked deliberately so the comparison can be viewed from two different angles: (a) a 1.3\,B-active MoE versus dense models with similar active count (active-parameter axis), and (b) a $\sim$4.2\,GB total-footprint MoE versus dense models with similar or smaller footprints (memory-footprint axis). Both readings get reported as reporting only one of them can be a misleading argument. On Jetson, we add a per-token energy measurement read directly from the SoC power rails via \texttt{tegrastats}.

We contribute the following:

\begin{enumerate}
  \item A direct, MELT-style head-to-head of MoE versus dense inference on a consumer laptop and an edge developer board. Throughput, peak memory, prompt-length sensitivity, and (on Jetson) joules per generated token are reported side by side under matched inference conditions.
  \item A reproducible measurement harness, including per-run JSONL records, the exact \texttt{llama-cli} command lines, a \texttt{tegrastats} parser, SHA-256 hashes of the GGUFs, and the hardware configuration and \texttt{llama.cpp} build flags pinned to a specific tag---released as an anonymized bundle.\footnote{\url{https://github.com/Analytics-Everywhere-Lab/edge-moe}}
  \item An evidence-based analysis of where MoE helps, where it does not, and which device characteristics dominate the outcome. The findings are bounded to two devices and four models, and we explicitly state the conditions under which we expect these findings to apply.
  \item A direct measurement of within-MoE routing-versus-FFN compute time on both backends, using a publicly available \texttt{llama.cpp} patch that aggregates per-node wall time via the \texttt{cb\_eval} scheduler hook. On Jetson (CUDA, 15\,W), routing is under 9\,\% of MoE-block decode time. On M2 (Metal), the same metric is affected by synchronization overhead but is capped at $\sim$33\,\%. Either reading rules out routing arithmetic as the bottleneck.
\end{enumerate}

This work introduces no new architecture; the contribution is a careful, reproducible measurement of an underexplored deployment setting, with code and per-run data released.

\section{Background and Related Work}
\label{sec:background}

A dense Transformer decoder layer applies the same set of weights to every input token, so the number of parameters that contribute to a single token's output is the model's total parameter count. An MoE layer replaces the single feed-forward network with $k$ parallel ``expert'' feed-forward networks plus a small router that picks, for each token, which $\text{top-}n < k$ experts to fire~\cite{shazeer2017outrageously,fedus2022switch}. Total parameter count grows with the number of experts, but active parameter count per token stays roughly constant. In FLOPs, cost tracks active parameters, while on disk and in memory, cost tracks total parameters.

That distinction matters more than it sounds on a device with a tight memory budget. Compute cost means an MoE with the same active-parameter count as a dense model should execute a comparable number of FLOPs per token. Memory cost means the full weight set has to live in memory regardless of which experts a particular token touches, because the router cannot predict the next token's assignment ahead of time. Bandwidth cost means MoE routing introduces irregular memory accesses that are friendly to large datacenter caches and less friendly to the small shared caches typical of edge SoCs. Whether sparse activation translates into a runtime advantage depends on which axis the device is bottlenecked on. On a compute-bound device, MoE should help. On a memory- or bandwidth-bound device, MoE may help less, not at all, or it may even cost more, because the full weight set still occupies the budget and the routing decisions add overhead. Which one applies on real hardware is an empirical question.

A growing line of work addresses the memory side of edge LLM inference. LLM-in-a-Flash~\cite{alizadeh2024flash} stores parameters in flash and streams them to DRAM on demand, motivated explicitly by the bandwidth gap between flash and DRAM. PowerInfer~\cite{song2023powerinfer} exploits activation sparsity to run dense LLMs on a single consumer GPU. For MoE specifically, Mixtral-offloading~\cite{eliseev2023fastmoe}, Fiddler~\cite{kamahori2024fiddler}, and MoE-Infinity~\cite{xue2024moeinfinity} all attack the dispatch-and-offload problem we surface in Section~\ref{sec:bottlenecks}, but they target multi-GPU servers or consumer GPUs with PCIe-attached CPU memory, not unified-memory edge SoCs. A recent survey~\cite{cai2026edge_survey} systematises the broader edge-LLM inference literature.

We standardise on \texttt{llama.cpp}~\cite{gerganov2024llamacpp}, a C/C++ inference engine designed to run LLMs with minimal setup across a wide range of hardware. It supports the Apple Metal backend on the M2 and the CUDA backend on the Jetson Ampere SoC, runs OLMoE (since PR \#9462, late 2024), and is widely used in the consumer and edge LLM space. Models are stored as GGUF. We use \texttt{Q4\_K\_M}, a 4-bit weight quantization that shrinks the model to roughly a quarter of its full-precision size, with finer-grained precision in the most sensitive tensors to limit the accuracy loss; it follows post-training weight-only methods like GPTQ~\cite{frantar2023gptq} and AWQ~\cite{lin2024awq}. It sits at a useful size-quality sweet spot and is consistently supported across all four models. We avoid \texttt{Q4\_0} for OLMoE because of documented assertion errors in community threads, and the \texttt{0125} OLMoE release because of similar known issues at the time of writing.

For methodology context, MELT~\cite{laskaridis2024melt} established that meaningful mobile-LLM measurement needs direct power instrumentation, sustained-performance testing, and explicit handling of thermal throttling, but covered dense models on phone-class devices. MoE-CAP~\cite{jiang2025moecap} established that meaningful MoE benchmarking needs sparsity-aware utilisation metrics, but covered datacenter GPUs. Closest in spirit to ours, Arya and Simmhan~\cite{arya2025edge} measure dense-LLM performance and power on the Jetson Orin AGX (64\,GB) across batch size, sequence length, and quantization, and we extend that line to MoE-versus-dense on a much tighter 8\,GB Jetson Orin Nano. Taken together, these threads leave a specific gap. Mobile-LLM measurement has covered dense models on phones~\cite{laskaridis2024melt}, MoE benchmarking has covered datacenter GPUs~\cite{jiang2025moecap}, edge-LLM measurement has covered dense models on developer boards~\cite{arya2025edge,husom2025sustainable,rajashekar2025sustainability}, and the MoE-offloading line targets multi-GPU or PCIe-attached settings rather than unified-memory edge SoCs~\cite{eliseev2023fastmoe,kamahori2024fiddler,xue2024moeinfinity}. To our knowledge, no prior work directly measures MoE-versus-dense inference under matched conditions on consumer and memory-constrained edge hardware. That gap is the reason this paper is empirical: the question of whether MoE's FLOP advantage survives on these devices is testable today, the answer has direct consequences for deployment decisions, and the existing literature does not answer it.

\section{Methodology}
\label{sec:methodology}

\subsection{Hardware}

Two devices, plus a development-only host that contributes no numbers (Table~\ref{tab:hardware}). The Jetson is run at the 15\,W power mode (\texttt{POWER\_MODEL ID=0} on this firmware, equivalent to \texttt{sudo nvpmodel -m 0}) with \texttt{sudo jetson\_clocks} enabled. This is the same mode MELT-style edge work uses, so our Jetson numbers are directly comparable, not the Super dev kit's higher MAXN\_SUPER ($\sim$25\,W) boost. The exact configuration (macOS / JetPack version, CPU and GPU core counts, storage, cooling, build flags, \texttt{llama.cpp} tag) is recorded in \texttt{HARDWARE.md} in the released bundle (see footnote~1). The M2 measurement session was run with non-essential apps quit, Spotlight indexing paused, and Time Machine off. The laptop was on AC the whole time.

\begin{table}[t]
\centering
\caption{Hardware roles. The DGX Spark is used only as a build environment and for smoke tests; it contributes no measurement to the paper.}\label{tab:hardware}
\setlength{\tabcolsep}{5pt}
\begin{tabularx}{\textwidth}{@{}l>{\raggedright\arraybackslash}X l>{\raggedright\arraybackslash}X@{}}
\toprule
Role & Device & Memory & Used for \\
\midrule
Consumer target  & Apple MacBook Pro M2 Pro      & 16\,GB unified  & all M2 measurements \\
Edge target      & NVIDIA Jetson Orin Nano 8\,GB & 8\,GB unified   & all Jetson measurements \\
Development only & NVIDIA DGX Spark              & 128\,GB unified & build environment and smoke tests only \\
\bottomrule
\end{tabularx}
\end{table}

\subsection{Backend}

A single \texttt{llama.cpp} tag on both devices (recorded in \texttt{HARDWARE.md}). On M2 we build with \texttt{-DGGML\_METAL=ON -DGGML\_METAL\_EMBED\_LIBRARY=ON}, while on Jetson with \texttt{-DGGML\_CUDA=ON -DCMAKE\_CUDA\_ARCHITECTURES=87}. We pass \texttt{-ngl 99} so every transformer layer is offloaded to the device's accelerator (Metal or CUDA). That removes a large source of variance in the load distribution between the CPU and GPU.

\subsection{Models}

Four GGUF models, all at \texttt{Q4\_K\_M} (Table~\ref{tab:models}). The three dense baselines are picked to cover both the active-parameter axis and the memory-footprint axis. Without both of these, this comparison would not be fair.

\begin{table}[t]
\centering
\caption{Models under test. ``Active'' and ``Total'' are in billions of parameters (B); ``Active'' is per-token and ``Total'' is on disk and in memory. Model references: OLMoE~\cite{muennighoff2024olmoe}, Llama-3.2~\cite{dubey2024llama3}, Qwen2.5~\cite{yang2024qwen25}, Gemma-2~\cite{gemmateam2024gemma2}.}\label{tab:models}
\begin{tabular}{@{}llrrr@{}}
\toprule
Role & Model & Active (B) & Total (B) & File (GB) \\
\midrule
MoE & OLMoE-1B-7B-0924-Instruct & 1.3 & 6.9 & 4.21 \\
Dense (active-match) & Llama-3.2-1B-Instruct & 1.0 & 1.0 & 0.81 \\
Dense (active-match) & Qwen2.5-1.5B-Instruct & 1.5 & 1.5 & 0.99 \\
Dense (memory anchor) & Gemma-2-2B-Instruct & 2.0 & 2.0 & 1.71 \\
\bottomrule
\end{tabular}
\end{table}

\subsection{Prompt set}

Twelve prompts spanning general QA, reasoning, coding, summarisation, instruction following, and long context. The full set is in \texttt{configs/prompts.yaml}. Long-context items are 1000--1500-word synthetic passages we wrote to avoid any copyright concern where the short and medium prompts are simple factual, reasoning, or coding tasks. Inference settings are fixed at \texttt{temperature=0.0}, \texttt{top\_p=1.0}, \texttt{seed=42}, and \texttt{max\_new\_tokens=256}.

\subsection{Metrics, repetitions, and reproducibility}

For each combination of device, model, prompt, and repetition, we log wall-clock time (Python-side, includes load), prompt-evaluation and generation tokens/sec (from \texttt{llama.cpp}'s perf footer), peak RSS (\texttt{psutil} polling at 0.2\,s, summed across the inference process tree), CPU\% mean and peak, and the GGUF file size. On Jetson we additionally record average and peak power during generation (\texttt{tegrastats} at $\sim$10\,Hz, sliced to the inference window), the integrated energy over that window (left-Riemann sum of the power samples), joules per generated token, and peak SoC temperature. Time alignment between \texttt{tegrastats} and the inference subprocess uses high-resolution Python timestamps captured immediately before and after each \texttt{llama-cli} call. Model loading and shutdown energy are excluded from the per-token figure.

Each (device, model, prompt) is repeated three times. The headline throughput, memory, and stability aggregates pool the twelve-prompt efficiency set with the three prompt-length prompts of Section~\ref{sec:promptlen}, for 45 runs per (device, model) cell. 

We report mean, median, standard deviation, and coefficient of variation across these runs. On Jetson we insert a 30-second cool-down between repetitions to keep thermal throttling out of the numbers. Peak SoC temperature is logged per run and no run crossed the 60\,$^{\circ}$C escape-hatch threshold (peak observed 55.5\,$^{\circ}$C on Gemma-2-2B), so the variance-inflation contingency (raise reps to five) was not triggered. The runner loads each GGUF once per model and reuses the OS page cache across its runs.

A thread-count sweep ($t \in \{4, 6, 8, 10\}$, 144 runs) put every model's optimum at $t = 4$. Our \texttt{-t 6} headline runs sit within 0.6--2.0\,\% of each per-model optimum and do not advantage either model class. The full sweep is in the bundle.

We deliberately report no M2 energy. macOS does not expose stable, well-documented per-rail power counters comparable to Jetson's SysFS rails. Tools like \texttt{powermetrics} report partly-inferred values that move with user-space workloads in ways that are not reviewer-defensible for a head-to-head MoE-versus-dense comparison. Reporting an M2 energy figure we cannot stand behind would be worse than reporting none. RQ4 is a Jetson question, and we are explicit about that scope.

For reproducibility, every JSONL line records the \texttt{git rev-parse HEAD}, the loaded GGUF's file size, and the exact \texttt{llama-cli} command string. SHA-256 hashes of all GGUFs land in \texttt{data/processed/model\_hashes.txt}. The \texttt{llama.cpp} tag and the unpatched and patched binary SHA-256 hashes are in \texttt{HARDWARE.md}.

\section{Results}
\label{sec:results}

\subsection{Throughput across devices and models (RQ1)}
\label{sec:throughput}

\begin{figure}[t]
\centering
\includegraphics[width=0.80\textwidth]{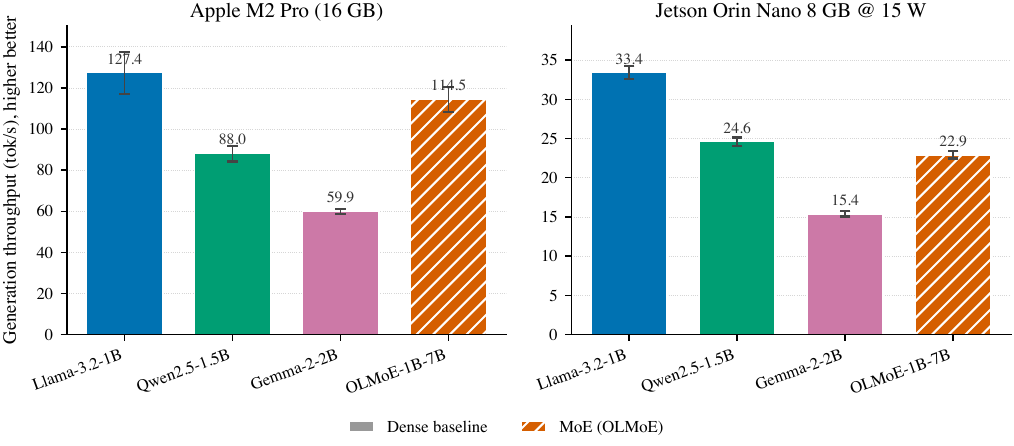}
\caption{Generation throughput by model and device. Error bars show $\pm 1$\,stdev across the 45 runs per cell (12 efficiency plus 3 sequence-length prompts, 3 reps each). OLMoE (hatched) lands between the two same-active dense baselines on both devices, and the gap to the fastest dense model widens from M2 to Jetson.}
\label{fig:throughput}
\end{figure}

For each pair of device and model we report mean and median generation tokens per second across all (prompt, repetition) combinations (Table~\ref{tab:throughput}, Fig.~\ref{fig:throughput}).

\begin{table}[t]
\centering
\caption{Generation throughput (tokens/sec). Mean and median across the 45 runs per cell (12 efficiency prompts plus 3 sequence-length prompts, 3 reps each). Source: \texttt{model\_overall\_summary.csv}.}\label{tab:throughput}
\begin{tabular}{@{}llrrrr@{}}
\toprule
Device & Model & Active (B) & Total (B) & Mean & Median \\
\midrule
M2 & Llama-3.2-1B & 1.0 & 1.0 & 127.4 & 130.7 \\
M2 & OLMoE-1B-7B & 1.3 & 6.9 & 114.5 & 116.2 \\
M2 & Qwen2.5-1.5B & 1.5 & 1.5 & 88.0 & 89.5 \\
M2 & Gemma-2-2B & 2.0 & 2.0 & 59.9 & 60.4 \\
\midrule
Jetson 15\,W & Llama-3.2-1B & 1.0 & 1.0 & 33.4 & 33.0 \\
Jetson 15\,W & Qwen2.5-1.5B & 1.5 & 1.5 & 24.6 & 24.5 \\
Jetson 15\,W & OLMoE-1B-7B & 1.3 & 6.9 & 22.9 & 23.0 \\
Jetson 15\,W & Gemma-2-2B & 2.0 & 2.0 & 15.4 & 15.4 \\
\bottomrule
\end{tabular}
\end{table}

The two comparison axes can be read off the same table. On the active-parameter axis, OLMoE sits between Llama-3.2-1B and Qwen2.5-1.5B on both devices, slower than Llama-3.2-1B (the smallest dense model with the smallest total weight) and faster than Qwen2.5-1.5B. The naive sparse-activation argument predicts OLMoE near Llama-3.2-1B; the data place it about 10\,\% below on M2 (114.5 vs 127.4\,tok/s) and about 31\,\% below on Jetson (22.9 vs 33.4\,tok/s). The active-parameter advantage is therefore only partially realised on the consumer device and substantially reduced on the edge device. On the memory-footprint axis, OLMoE outpaces Gemma-2-2B on both devices (114.5 vs 59.9\,tok/s on M2, 22.9 vs 15.4\,tok/s on Jetson) at the cost of a $\sim$2.5$\times$ larger memory footprint (file-size basis: 4.21 vs 1.71\,GB). So a fixed-budget edge deployment that swaps Gemma-2-2B for OLMoE pays a memory tax for a throughput gain.

The qualitative ordering is the same on both devices, but the gap between MoE and the fastest dense baseline widens from M2 to Jetson. That is consistent with the memory-bandwidth account in Section~\ref{sec:bottlenecks} where the narrower bandwidth on the Jetson highlights the overhead associated with loading the entire set of weights, while routing arithmetic itself is not the issue.

\subsection{Peak memory across devices and models (RQ2)}
\label{sec:memory}

\begin{figure}[t]
\centering
\includegraphics[width=0.80\textwidth]{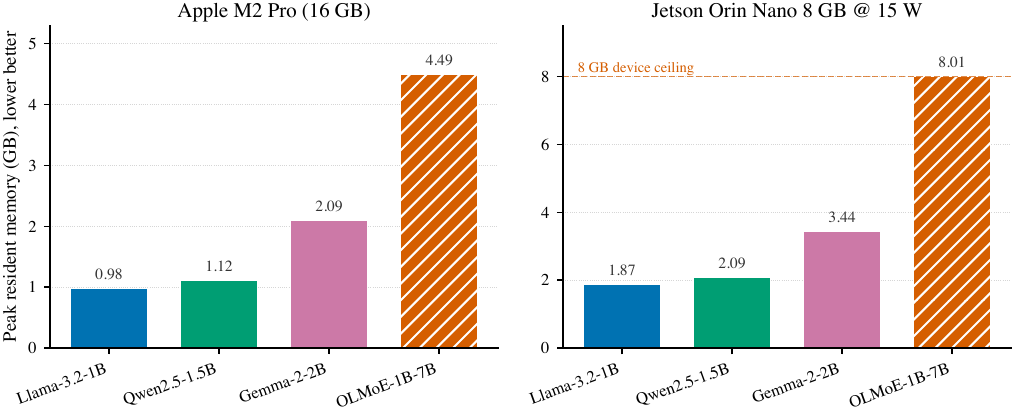}
\caption{Peak resident memory by model and device. OLMoE (hatched) sits right at the Jetson's 8\,GB physical ceiling and is the only model in our set that does.}
\label{fig:memory}
\end{figure}

Peak resident memory tracks total parameter count, not active parameter count (Table~\ref{tab:memory}, Fig.~\ref{fig:memory}). OLMoE consumes 4.5\,GB on M2 and 8.0\,GB on Jetson. The Jetson figure is right at the device's 8\,GB physical ceiling and survives only because the kernel-default zram swap absorbs the excess data. A dense model with OLMoE's 1.3\,B \emph{active} parameters would fit in 0.8--1.0\,GB. OLMoE's footprint is 5--10$\times$ larger than that.

\begin{table}[t]
\centering
\caption{Peak resident memory by device and model. File size is in decimal GB ($10^9$ bytes); peak RSS is in binary GiB ($2^{30}$ bytes), which is the unit the 8\,GB Jetson ceiling is specified in. ``Peak/file ratio'' is taken as displayed; correcting for the unit mismatch lowers each ratio by $\sim$7\,\%. Source: \texttt{model\_overall\_summary.csv}, \texttt{cross\_device\_comparison.csv}.}\label{tab:memory}
\begin{tabular}{@{}llrrrr@{}}
\toprule
Device & Model & Total (B) & File (GB) & Peak RSS (GiB) & Peak/file \\
\midrule
M2 & Llama-3.2-1B & 1.0 & 0.81 & 0.98 & 1.21 \\
M2 & Qwen2.5-1.5B & 1.5 & 0.99 & 1.12 & 1.13 \\
M2 & Gemma-2-2B & 2.0 & 1.71 & 2.09 & 1.22 \\
M2 & OLMoE-1B-7B & 6.9 & 4.21 & 4.49 & 1.07 \\
\midrule
Jetson 15\,W & Llama-3.2-1B & 1.0 & 0.81 & 1.87 & 2.31 \\
Jetson 15\,W & Qwen2.5-1.5B & 1.5 & 0.99 & 2.09 & 2.11 \\
Jetson 15\,W & Gemma-2-2B & 2.0 & 1.71 & 3.44 & 2.01 \\
Jetson 15\,W & OLMoE-1B-7B & 6.9 & 4.21 & 8.01 & 1.90 \\
\bottomrule
\end{tabular}
\end{table}

The Jetson peak/file ratio (around 2.0$\times$) is consistently higher than the M2 ratio (around 1.1--1.2$\times$). The CUDA path needs a separate device-side allocation on top of the host-side mmap; M2's unified memory has no such duplication. On a fixed 8\,GB Jetson budget, OLMoE is the only model in our set that approaches the upper limit.

\subsection{Prompt-length sensitivity (RQ3)}
\label{sec:promptlen}

We hold generated tokens fixed at 256 and vary prompt length across short ($\sim$50--100 words), medium ($\sim$300--500 words), and long ($\sim$1000--1500 words). Any change in generation throughput is therefore caused by prompt length, not by a difference in generation length (Table~\ref{tab:promptlen}).

\begin{table}[t]
\centering
\caption{Prompt-length sensitivity. Generation tok/s, mean across three prompts $\times$ three reps per length-bucket. $\Delta$ is (long $-$ short) / short. Source: \texttt{sequence\_length\_summary.csv}.}\label{tab:promptlen}
\begin{tabular}{@{}llrrrr@{}}
\toprule
Device & Model & Short & Medium & Long & $\Delta$ \\
\midrule
M2 & Llama-3.2-1B & 134.4 & 130.2 & 107.8 & $-$19.7\,\% \\
M2 & OLMoE-1B-7B & 119.2 & 115.0 & 103.8 & $-$12.9\,\% \\
M2 & Qwen2.5-1.5B & 90.6 & 89.2 & 80.7 & $-$10.9\,\% \\
M2 & Gemma-2-2B & 60.7 & 60.3 & 57.4 & $-$5.4\,\% \\
\midrule
Jetson 15\,W & Llama-3.2-1B & 34.3 & 32.8 & 32.8 & $-$4.4\,\% \\
Jetson 15\,W & Qwen2.5-1.5B & 25.1 & 24.4 & 23.8 & $-$5.2\,\% \\
Jetson 15\,W & OLMoE-1B-7B & 23.3 & 22.9 & 22.2 & $-$4.7\,\% \\
Jetson 15\,W & Gemma-2-2B & 15.7 & 15.4 & 14.8 & $-$5.4\,\% \\
\bottomrule
\end{tabular}
\end{table}

Two things stand out. First, OLMoE does not degrade more than the dense baselines as the prompt length grows. Its slope on each device is comparable to that of the corresponding dense model. The theoretical expectation that an MoE's per-token routing cost should make it more sensitive to prompt length is not supported at this parameter scale. Second, the long-prompt OLMoE runs on Jetson all completed without OOM at peak RSS 8.01\,GB, slightly above the 8\,GB physical ceiling, absorbed by the kernel-default zram swap. OOM remains a real risk for any deployment that disables swap or runs another resident process, and we report this as a finding rather than as a successful completion.

\subsection{Stability across repetitions}
\label{sec:stability}

Across the 45 runs per (device, model) cell, Jetson is tight (CV 2.0--2.4\,\%) and three of four M2 models sit at CV $\leq$ 5.2\,\%. Llama-3.2-1B is the outlier at \textbf{8.1\,\%}, holding even after a clean re-run; this is intrinsic to the smallest, fastest model, at $\sim$127\,tok/s a 256-token run completes in $\sim$2\,s, so a single scheduler stall swings a run by $\sim$5\,\%, rather than background contention. It stays below our 10\,\% threshold and we flag it as a known limitation of the M2 stack at this model scale.

\subsection{Energy per generated token on Jetson (RQ4)}
\label{sec:energy}

\begin{figure}[t]
\centering
\includegraphics[width=0.72\textwidth]{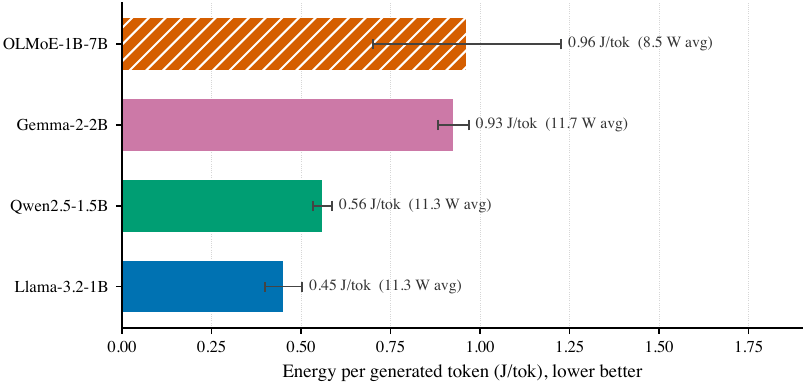}
\caption{Jetson energy per generated token at the 15\,W envelope, sorted ascending. Annotations show the mean joules per token and the average wall power during generation. OLMoE costs about 2.1$\times$ as much as Llama-3.2-1B for a comparable active count, while running at a lower duty-cycle wall power (8.5\,W).}
\label{fig:energy}
\end{figure}

We integrate \texttt{tegrastats} SoC-rail power over the inference window and divide by generated-token count (Table~\ref{tab:energy}, Fig.~\ref{fig:energy}).

\begin{table}[t]
\centering
\caption{Jetson 15\,W energy per generated token. Source: \texttt{jetson\_energy\_summary.csv}.}\label{tab:energy}
\begin{tabular}{@{}lrrrrr@{}}
\toprule
Model & Mean J/tok & Median J/tok & Avg power (W) & Peak power (W) & Peak SoC $^{\circ}$C \\
\midrule
Llama-3.2-1B & \textbf{0.45} & 0.44 & 11.31 & 12.89 & 53.2 \\
Qwen2.5-1.5B & 0.56 & 0.56 & 11.32 & 12.30 & 53.3 \\
Gemma-2-2B & 0.93 & 0.91 & 11.67 & 12.74 & 55.5 \\
OLMoE-1B-7B & 0.96 & 0.86 & 8.49 & 10.97 & 55.1 \\
\bottomrule
\end{tabular}
\end{table}

Across our Jetson runs, OLMoE costs about 2.1$\times$ as much energy per token as Llama-3.2-1B, even though the two models have comparable active-parameter counts (1.3\,B active vs 1.0\,B dense). This increase in energy consumption is exactly what the peak memory usage data in Section~\ref{sec:memory} predict. Fetching expert weights from the full 6.9\,B-parameter set at every MoE layer consumes more memory-bandwidth energy than fetching the smaller 1.0\,B-parameter dense model, even though the floating-point operation count per token is similar. So, energy on edge tracks total parameters, not active parameters. Our per-token figures sit in the same range as recent dense-LLM edge-energy measurements~\cite{husom2025sustainable,rajashekar2025sustainability}, none of which cover MoE.

OLMoE also runs at a notably lower average power consumption (8.5\,W) than the dense baselines (11.3--11.7\,W). That is consistent with longer per-token inference times (lower duty cycles on the compute units), compounded by memory-bandwidth-bound execution. The energy-per-token figure is high because total inference time is long, not because instantaneous power is high. None of the models came near the 15\,W envelope's thermal headroom. Peak SoC temperature stayed in the 53--56\,$^{\circ}$C range, well below the throttle threshold.

\subsection{Router-overhead breakdown}
\label{sec:router}

To investigate what is actually slowing OLMoE down, we patched a copy of \texttt{llama.cpp} \texttt{b4404} to install an evaluation callback on the ggml scheduler that observes every node in the decode graph. Each node's wall-clock compute time gets categorized into ``routing/gating'' or ``expert FFN'' using the existing \texttt{ffn\_moe\_*} tag set generated by \texttt{llm\_build\_moe\_ffn} in \texttt{src/llama.cpp}. The patch is a public-API-only addition to \texttt{examples/main/main.cpp} (no internal changes). It is saved as \texttt{scripts/patches/router\_overhead\_b4404.patch}. The patched binary is used only for the runs in this subsection; the headline numbers in Sections~\ref{sec:throughput}--\ref{sec:energy} all come from the unpatched binary, with separate \texttt{m2\_patched} and \texttt{jetson\_patched} device profiles in \texttt{configs/devices.yaml} and binary SHA-256 hashes in \texttt{HARDWARE.md}.

Observing every node forces the scheduler to synchronize between nodes, so the patched binary runs OLMoE substantially slower than the unpatched binary by roughly 15$\times$ on Metal and only $\sim$1.3$\times$ on CUDA, which is an asymmetry we quantify below and rely on for the rest of this subsection. This slowdown does not affect routing and FFN nodes equally. The per-node synchronization adds a roughly fixed cost per graph node, large relative to the true compute of a cheap routing node (softmax + top-k) but small relative to that of an expensive expert-FFN node. The measured routing share is therefore biased \emph{upward} by the instrumentation, and the size of that bias depends on how expensive per-node sync is on a given backend. We do not claim the bias cancels in the ratio; instead, we use the cross-backend comparison to bound it. Because CUDA's per-node sync is far cheaper than Metal's (quantified below), the Jetson measurement carries far less of this inflation and is the trustworthy estimate, while the M2 measurement is an upper bound.

\paragraph{The metric we report.} The clean, paper-defensible quantity is the within-MoE-block routing-to-FFN time ratio:
\[
\text{routing share} = \frac{t_\text{routing}}{t_\text{routing} + t_\text{expert FFN}}.
\]
``Fraction of total decode time in MoE blocks'' is \emph{not} a clean metric on the patched binary, because non-MoE nodes (attention, embeddings, sampling) outnumber MoE nodes roughly 4:1 in the OLMoE graph. The per-node sync inflates the non-MoE share disproportionately, for the same reason it inflates routing: attention, embedding, and sampling nodes are numerous and individually cheap, so a fixed per-node cost weighs heavily on them. We therefore report the within-MoE split and refrain from making a ``\% of decode'' claim that the instrument cannot support.

\paragraph{Results.} Across 45 patched OLMoE runs (12 prompts $\times$ 3 reps) on each backend, the routing share is bimodal in prompt length. On M2 (Metal) it clusters at $\sim$33.5\,\% for decode-dominated workloads (prompts $\leq$ 200 tokens) and $\sim$10.2\,\% for prefill-heavy workloads ($>$ 500 tokens). On Jetson (CUDA, 15\,W) the same protocol gives much lower numbers on both strata with $\sim$8.9\,\% decode-dominated and $\sim$7.3\,\% prefill-heavy. Routing-node absolute time stays roughly constant within each backend ($\sim$2.9--3.1\,s on M2, $\sim$0.6--0.7\,s on Jetson) while expert-FFN time scales sharply with prompt length (M2: 5.6\,s $\rightarrow$ 27.8\,s, Jetson: 6.5\,s $\rightarrow$ 8.7\,s). That is the expected architectural shape where routing is $O(\text{top}_k \cdot n_\text{experts})$ and dominated by softmax + top-$k$, which does not scale with prompt length, while expert-FFN is $O(N_\text{tokens} \cdot \text{top}_k \cdot d^2)$ and scales with how many tokens are in flight.

\begin{table}[t]
\centering
\caption{Router overhead within OLMoE MoE blocks, by backend and prompt stratum. The pooled ``all 12$\times$3'' rows are bimodal in prompt length, so we report the two strata directly. The routing-share column is the mean of per-run within-MoE shares; dividing the averaged Routing and FFN columns gives a close but slightly different value. Source: \texttt{router\_overhead\_breakdown.csv}.}\label{tab:router}
\setlength{\tabcolsep}{4pt}
\small
\begin{tabular}{@{}llrrrrr@{}}
\toprule
Device & Stratum & $n$ & Routing & FFN & MoE & Routing \\
       &         &     & (ms)    & (ms) & (ms) & share \\
\midrule
M2 (Metal)   & decode-dom.\ ($\leq$200) & 30 & 2871 & 5587 & 8458 & \textbf{33.5\,\%} \\
M2 (Metal)   & prefill-heavy ($>$500)   & 6  & 3059 & 27751 & 30810 & 10.2\,\% \\
Jetson 15\,W & decode-dom.\ ($\leq$200) & 30 & 635  & 6534 & 7169 & \textbf{8.9\,\%} \\
Jetson 15\,W & prefill-heavy ($>$500)   & 6  & 686  & 8712 & 9398 & 7.3\,\% \\
\bottomrule
\end{tabular}
\end{table}

\paragraph{Backend asymmetry: why M2 and Jetson disagree.} The decode-dominated routing-share delta is $33.5 - 8.9 = 24.6$\,pp, well above the 15\,pp threshold at which a naive cross-backend comparison would warrant caution. The cause, however, is identifiable from the patched-binary slowdown itself rather than from any architectural difference in the routing-versus-FFN compute mix:

\begin{itemize}
  \item Patched-binary slowdown on M2 (Metal): OLMoE drops from 114.5 unpatched gen tok/s to 7.7 patched = \textbf{14.9$\times$ slowdown} under \texttt{cb\_eval} per-node sync.
  \item Patched-binary slowdown on Jetson (CUDA, 15\,W): OLMoE drops from 22.9 unpatched gen tok/s to 18.2 patched = \textbf{1.26$\times$ slowdown} under the same per-node sync.
\end{itemize}

CUDA's per-node \texttt{ggml\_backend\_synchronize} is roughly 12$\times$ cheaper than Metal's. Combined with OLMoE having more routing nodes (6240) than FFN nodes (5200), Metal's heavier sync lands disproportionately on routing and inflates the apparent M2 routing share; CUDA's cheap sync leaves the Jetson share close to the true within-MoE compute split.

\begin{figure}[tp]
\centering
\includegraphics[width=0.72\textwidth]{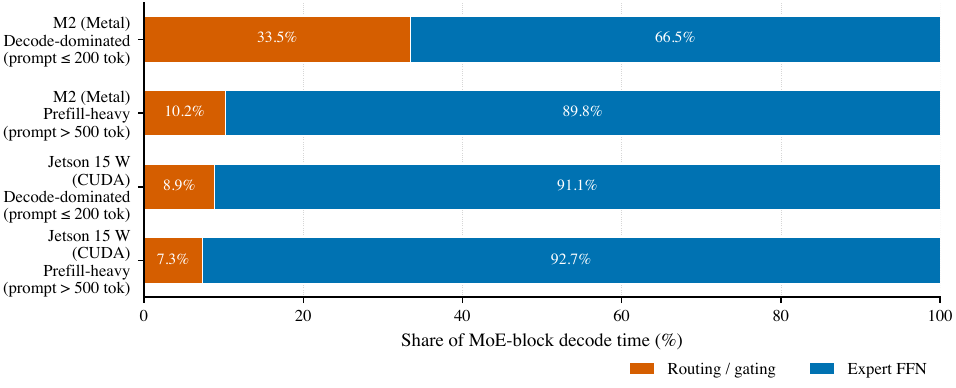}
\caption{Within-MoE routing-versus-FFN time, by backend and prompt stratum. Expert FFN dominates everywhere. The M2 decode-dominated 33.5\,\% routing share is inflated by Metal's heavier per-node \texttt{cb\_eval} sync overhead and should be read as an upper bound; the Jetson 8.9\,\% is the cleaner estimate.}
\label{fig:router}
\end{figure}

\section{Discussion}
\label{sec:discussion}

\subsection{Does MoE help on consumer hardware?}

On the M2 Pro the active-parameter advantage shows up only partially: OLMoE is faster than the memory-anchor baseline (Gemma-2-2B) but slower than the same-active-count neighbour (Llama-3.2-1B), which has nearly 7$\times$ fewer total parameters (1.0\,B vs 6.9\,B). Whether that counts as ``MoE helping'' depends on whether quality per token, which we do not measure, repays the 4.5\,GB memory tax. On a 16\,GB unified-memory laptop, the tax is comfortable and the answer is ``yes, modestly.'' On a narrower 8\,GB laptop, it becomes a hard ceiling.

\subsection{Does MoE help on edge hardware?}
On the Jetson the active-parameter advantage is fully eroded. OLMoE is slower than every dense baseline except Gemma-2-2B, the throughput gap to the same-active Llama-3.2-1B widens from 10\,\% on M2 to 31\,\%, and energy per token is 2.1$\times$ higher, consistent with the bandwidth account in Section~\ref{sec:bottlenecks}. It additionally peaks at the device's 8\,GB ceiling and fits only because zram swap absorbs the spill, where all three dense baselines fit comfortably (1.87--3.44\,GB).

So under the evaluated conditions the answer is ``no, not at this parameter scale'': throughput is lower, energy per token is higher, and the memory footprint sits at the hard ceiling. This contradicts, on direct measurement, the intuition that ``MoE should be ideal for edge because it computes less per token,'' at least for the OLMoE model on 8\,GB Jetson hardware.

\subsection{Where the bottleneck actually sits}
\label{sec:bottlenecks}

OLMoE's peak memory is roughly equal to its on-disk file size on both devices, so total parameter count dominates the memory footprint regardless of which 1.3\,B subset is active per token. Sparse activation does not buy back memory at this parameter scale.

For throughput, the router-overhead measurement in Section~\ref{sec:router} provides specific numbers for both backends. On Jetson (CUDA, 15\,W), routing/gating accounts for $\sim$8.9\,\%  of MoE-block compute in decode-dominated workloads (chat-style: short prompt, long generation) and $\sim$7.3\,\% in prefill-heavy ones (RAG-style: long context, short generation). This is the cleaner of the two measurements because CUDA's per-node sync is roughly 12$\times$ cheaper than Metal's, so it contaminates the timing far less. On M2 (Metal), the same protocol reports a higher share, $\sim$33.5\,\% decode-dominated and $\sim$10.2\,\% prefill-heavy, but a large fraction of the M2 number is per-node sync overhead, inflating routing-node accounting. The M2 figure is best read as an upper bound on the true within-MoE routing share.

Either reading rejects the naive ``MoE is slow on edge because routing arithmetic is expensive'' hypothesis. On the cleaner Jetson measurement, routing is under 9\,\% of MoE compute in decode-dominated workloads and under 8\,\% in prefill-heavy ones. Even on the sync-inflated M2 measurement, routing peaks at 1/3 of MoE compute, with expert FFN dominating by roughly 2:1.

If routing is not the within-MoE bottleneck, the throughput gap to dense baselines of comparable active-parameter count has to come from somewhere outside the routing arithmetic itself:

\begin{enumerate}
  \item \textbf{Total parameter footprint and memory bandwidth.} OLMoE's 6.9\,B total parameters (vs Llama-3.2-1B's 1.0\,B) require the full expert weights to be resident on each layer where they may be consulted, even when only 8 of 64 experts fire per token. The per-token bandwidth requirement is a function of the total expert weight that must be paged through the compute units, not just the active parameter count.
  \item \textbf{Per-token expert dispatch.} Each layer's MoE FFN selects 8 of 64 experts per token. The dispatch-and-gather pattern (top-$k$ indexing, gather, scatter, weighted aggregation) is structural to MoE inference and absent from dense models, regardless of how cheap routing arithmetic is. Our measurement times the arithmetic, not the dispatch bookkeeping (allocator path, scheduler launch), folded into ``other'' in Section~\ref{sec:router}.
  \item \textbf{KV-cache pressure on long prompts.} KV-cache size grows with prompt length and model depth (16 layers in OLMoE), identically between dense and MoE at the same depth. This is not MoE-specific, but it interacts with the bandwidth point on memory-constrained edge hardware. Server-side, KV-cache fragmentation is the problem that PagedAttention~\cite{kwon2023vllm} addresses; on the edge, the constraint is harder because there is no separate host memory to page into. This shows up directly in our data, where on Jetson long-prompt OLMoE runs hug the 8\,GB ceiling (Section~\ref{sec:promptlen}).
\end{enumerate}

Routing share differs by 24.6\,pp between M2 and Jetson in the decode-dominated stratum. The patched-binary slowdown asymmetry (M2 14.9$\times$ vs Jetson 1.26$\times$, Section~\ref{sec:router}) points to backend-specific \texttt{cb\_eval} sync overhead as the cause rather than to an architectural difference between routing and FFN compute. We therefore report Jetson's 8.9\,\% as the primary number, M2's 33.5\,\% as a sync-inflated upper bound, and we do not make a within-MoE compute claim on M2 alone.

\subsection{Connection to MoE-CAP's S-MBU and S-MFU}

MoE-CAP~\cite{jiang2025moecap} proposes Sparsity-Aware Memory Bandwidth Utilisation (S-MBU) and Sparsity-Aware Model FLOPs Utilisation (S-MFU) for characterising MoE efficiency in datacenter conditions. Our wall-clock tokens/sec, peak RAM, and joules per token can be read as device-specific projections of the same underlying quantities. A low S-MBU prediction would manifest, in our setting, as wall-clock throughput materially below the active-parameter expectation, which is exactly what we see on Jetson. We use S-MBU/S-MFU as an analytical lens for our measured numbers rather than reproduce them directly, since they require backend-internal counters that \texttt{llama.cpp} does not expose.

\subsection{Implications for hardware-aware MoE design}
The data motivate three concrete directions for OLMoE-class edge-MoE deployment. First, routing arithmetic itself is cheap (under 9\,\% of MoE-block decode on Jetson CUDA), so what actually matters is which experts get selected, as a cache-aware routing strategy could pull the energy penalty back closer to active-parameter cost. Second, OLMoE \texttt{Q4\_K\_M} sits right at the 8\,GB Jetson ceiling, so one extra quantization step on the expert weights (\texttt{Q4\_K\_S}, \texttt{Q3\_K\_M}, possibly asymmetric over less-frequently-activated experts) would buy several hundred MB of headroom and could unlock long-prompt deployment without zram-swap dependence. Third, per-token slowdown is comparable between MoE and dense, but absolute memory pressure is higher, so KV-cache-aware serving~\cite{kwon2023vllm} matters more on edge MoE than dense at the same active count.
  
\section{Threats to Validity}
\label{sec:threats}

\textbf{Hardware coverage.} One consumer-class device (M2 Pro) and one edge-class device (Jetson Orin Nano). Findings should not be extrapolated to phone-class silicon (Snapdragon, Apple A-series), other consumer GPUs, or other edge SoCs, without further measurement.

\textbf{One MoE, three dense baselines.} OLMoE-1B-7B is one point in MoE architecture space, so conclusions about routing overhead, expert sizes, and total/active ratios are tied to its design and may not transfer to Mixtral-class~\cite{jiang2024mixtral}, DeepSeek-MoE-class~\cite{dai2024deepseekmoe}, or Qwen-MoE-class models. The dense baselines were picked to span the active-parameter and memory-footprint axes, and other picks could shift the quantitative numbers.

\textbf{Backend and prompt set.} The numbers are tied to one \texttt{llama.cpp} tag. vLLM, MLC-LLM, or TensorRT-LLM would likely give different absolute numbers, especially on Jetson. Our prompt set has twelve items, the longest at 1500 words. We do not stress-test multi-thousand-token prompts or multi-turn conversations.

\textbf{Energy measurement only on Jetson.} macOS power tooling is too noisy for a reviewer-defensible cross-model comparison, so we report no M2 energy. The asymmetry is deliberate. OLMoE \texttt{Q4\_K\_M} may OOM on long prompts on the 8\,GB Jetson, and we report that as a finding rather than immediately performing re-quantization. A Snapdragon or A-series phone measurement is the obvious next measurement and is explicitly out of scope here.

\section{Conclusion and Future Work}
\label{sec:conclusion}
We conclude that, for the OLMoE model we evaluate at this parameter scale, sparse activation does not deliver the edge-inference advantage it promises: \emph{on bandwidth-bound hardware, inference cost tracks total parameters, not active ones}.

Using direct measurement, OLMoE-1B-7B's active-parameter advantage over similarly-active dense baselines is only partially realised on the M2 Pro ($\sim$10\,\% below Llama-3.2-1B at matched active count) and erodes on the Jetson Orin Nano (15\,W, $\sim$31\,\% below, at 2.1$\times$ the energy per token and peak RSS at the 8\,GB ceiling). A patched-binary measurement of the within-MoE compute split rules out routing arithmetic as the bottleneck: the gap is memory-bandwidth-driven, the one property of OLMoE that does not benefit from sparse activation. Replacing a comparably active dense baseline with an MoE on edge hardware at this scale therefore tends to raise memory use and lower throughput and energy efficiency. The open question for future MoE design is whether sparse-activation quality gains, with hardware-aware routing, can offset that overhead.

Future work opportunities fall into three primary areas. The first is phone-class measurement, extending MELT's methodology~\cite{laskaridis2024melt} to MoE with PowerInfer-2-style~\cite{xue2024powerinfer2} deployment on Snapdragon and A-series silicon, the next step after the datacenter-laptop-board progression. The second is broader MoE coverage (Qwen3-MoE, DeepSeek-V2-Lite-MoE, Mixtral-class) at edge-tractable quantizations, against mobile-optimised dense baselines like MobileLLM~\cite{liu2024mobilellm}. The third is cache-aware routing that accounts for which experts are resident versus on disk. The numbers we report motivate all three; designing and validating the strategies is separate work.

\section*{Acknowledgment}
This work was supported by NSERC Discovery Grant No RGPIN-2025-04478 and NSERC Discovery Supplement Award No DGECR-2025-00129.



\bibliographystyle{splncs04}
\bibliography{references}

\end{document}